\documentclass[12pt,preprint]{aastex}
\def\,{\thinspace}
\def\LprimeCO{{\hbox {$L^\prime_{\rm CO}$}}}
\def\Lprimeco{{\hbox {$L^\prime_{\rm CO}$}}}
\def\Msun{M$_\odot$}
\def\Lsun{L$_\odot$}

\def \kms{km\,s$^{-1}$}
\def \Kkmspc{K\,\kms\,pc$^2$}
\received{28 June 2002}
\slugcomment{}
\begin{document}
\title{
MOLECULAR GAS AND DUST AT $z$ = 2.6 IN SMM J14011+0252:
A STRONGLY LENSED, ULTRALUMINOUS GALAXY, NOT A HUGE, MASSIVE DISK.}
\author{D.~Downes}
\affil{Institut de Radio Astronomie Millim\'etrique, Domaine Universitaire,
           F-38406 St. Martin d'H\`eres,  France}
\author{P.M.~Solomon}
\affil{
Astronomy Program, State University of New York, Stony Brook, NY 11794
}
\begin{abstract}
We used the IRAM Interferometer to detect CO(3--2), CO(7--6), and
1.3\,mm dust continuum emission from the submillimeter galaxy SMM
J14011+0252 at $z=2.6$.  Contrary to a recent claim that the CO was
extended over 6.6$''$ (57\,kpc), the new data yield a size of $2''\times \leq
0.5''$ for the CO and the dust.  Although previous results placed the
CO peak in a region with no visible counterpart, the new maps show the
CO and dust are centered on the J1 complex seen on {\it
K}-band and optical images.  We suggest the CO is gravitationally
lensed not only by the foreground cluster A1835, but also by an
individual galaxy on the line of sight.  Comparison of measured and
intrinsic CO brightness temperatures indicates the CO size is
magnified by a factor of $25\pm 5$.  After correcting for lensing, we
derive a true CO diameter of $\sim 0.08''$ (700\,pc), consistent with
a compact circumnuclear disk of warm molecular gas similar to that in
Arp~220.  The high magnification means the true size, far-IR
luminosity, star formation rate, CO luminosity, and molecular gas mass
are all comparable with those in present-epoch ultraluminous IR galaxies, 
not with a huge, massive, early-universe galactic disk.
\end{abstract}


\section{EVIDENCE FOR A STARBURST IN SMM J14011+0252}
In the past decade, about twenty CO sources at
redshifts from $z = 2$ to 4.7 have been detected at millimeter
wavelengths.  Most of these high-$z$ molecular-line emitters are
{\it gravitationally lensed}, ultraluminous infrared galaxies
(ULIGs), often hosting quasars and vigorous starbursts that may
dominate star formation in the early universe. One of these
galaxies is SMM J14011+0252 at $z = 2.565$, originally found by
Smail et al.\ (1998).  It is located behind the galaxy cluster
Abell 1835 at $z=0.25$, and is relatively intense in the
submillimeter band, with an 850\,$\mu$m flux density of $15\pm
2$\,mJy (Ivison et al.\ 2000).  The visible spectrum of
SMM~J14011+0252, taken with the Keck-II telescope, covers
rest-frame wavelengths 1200--2400\,\AA , and has a blue continuum
with weak, narrow Ly$\alpha$ emission and UV absorption lines of
C\,{\sc iv}, Si\,{\sc ii}/ O\,{\sc i}, Al\,{\sc iii}, Si\,{\sc
iv}, and C\,{\sc ii}, all typical of star-forming galaxies (Ivison
et al.\ 2000).  There is no sign of highly-ionized species like
N\,{\sc v} and no broad Ly$\alpha$ or H$\alpha$ features, so there
is no evidence for emission from a central black hole accretion
disk.  The H$\alpha$-to-[N~II] ratio is 3, as usually found in
H~II regions. The 21.5\,cm radio flux of $\sim 100$\,$\mu$Jy
agrees with the far-IR-radio correlation for nearby starburst
galaxies (Ivison et al.\ 2000; 2001).  All these facts indicate
14011+0252 is powered by star formation rather than a
dust-shrouded quasar.

The CO(3--2) line in SMM J14011+0252 was first detected with the
Caltech interferometer (Frayer et al.\ 1999).  New results by Ivison
et al. (2001) seemed to show the CO was extended over 6.6$''$, or
57\,kpc.  Even allowing for the magnification of 2.5 assumed by Ivison
et al., the CO scale length would be $>20$\,kpc, more than twenty
times larger than the CO-emitting, starburst nuclear disks in local
ultraluminous IR galaxies.  The huge size and CO luminosity lead to a
picture of an immense protogalaxy with $\geq 10^{11}$\,\Msun \ of
metal-enriched, {\it molecular} gas (Ivison et al.\ 2001).  This is
equal to or greater than the stellar mass of all modern spiral
galaxies and is surpassed only by the baryonic mass of giant
ellipticals.  The existence of such a huge and massive system at $z$ =
2.6 would have profound implications for theories of galaxy formation.

Ivison et al.\ also measured a CO centroid that did not coincide with
the optical or {\it K}-band objects seen with the {\it Hubble Space
Telescope (HST)} or the United Kingdom Infrared Telescope (UKIRT),
thus implying that most of the large starburst region is completely
hidden from view. To verify the CO extent and position, we re-observed
the source at 1.3\,mm and 3\,mm with the IRAM Interferometer on
Plateau de Bure, France.  We find a much smaller source, centered on a
{\it K}-band feature, that we believe is strongly gravitationally
lensed, leading to a picture of SMM J14011+0252 as an early-universe
ultraluminous galaxy with a compact star-forming disk in its center.

\section{DETECTIONS, POSITIONS, AND SIZES}
\subsection{\it Observing Technique}
The interferometer recorded data at 1.3\,mm and 3\,mm simultaneously,
with six 15-m antennas spaced from 24\,m to 330\,m. The most compact
array, with spacings $\leq 80$\,m, integrated for 24 hours in
excellent weather with $\sim$2\,mm of precipitable water vapor and
r.m.s.\ phase errors of $\leq 20^\circ$ at 1.3\,mm.  The
longer-baseline configurations had another 16 hours with 1.3\,mm phase
errors $\leq 40^\circ$. The SIS receivers had equivalent system
temperatures outside the atmosphere of 150\,K at 3\,mm (97\,GHz) in
the lower sideband, and 250 to 400\,K at 1.3\,mm in upper and lower
sidebands separated by 3\,GHz, with the upper band at 226\,GHz.  The
spectral correlators covered 1700\,\kms \ at 3\,mm and 800\,\kms \ at
1.3\,mm, with resolutions of 8 and 4\,\kms , respectively. The
amplitude calibrators were 3C273 (18.2\,Jy at 3\,mm and 12.9\,Jy at
1.3\,mm), and MWC349 (1.0 and 1.7\,Jy at 3 and 1.3\,mm). The
uncertainties in the flux scales are $\pm 5$\% at 3\,mm and $\pm 10$\%
at 1.3\,mm.

The observing program monitored phases every 20\,min, at 3 and 1.3\,mm
simultaneously, on the nearby calibrators IAP 1413+135 and 1334$-$127,
which are northeast and southwest of 14011+0252  (Table~1).  The
data recording routine uses the 1.3\,mm total power to correct
amplitudes and phases at 3 and 1.3\,mm for short-term changes in
atmospheric water vapor.  A subsequent calibration program scales the
3\,mm curve of phase vs.\ time to 1.3\,mm, subtracts it from the
observed 1.3\,mm calibrator phases, and fits the phase difference
between the two receivers.  To all the visibilities, the calibration
program assigns weights proportional to the integration time and the
inverse square of system temperature.  One can then make maps either
with these weights (``natural weighting'') or with uniform weighting
of the $u,v$ data.

\subsection{\it CO Lines at 3 and 1.3\,mm}
We detected both CO(3--2) and CO(7--6) with good sensitivity 
(Figs.\,1---3).  The CO (3--2) flux, linewidth, and redshift all
agree with the results of Frayer et al.\ (1999). The apparent CO(3--2)
luminosity is $1\times 10^{11}$\,K\,\kms\,pc$^2$, a very high value,
well beyond the maximum of the CO power histogram for ultraluminous
galaxies. For comparison, the galaxy Arp\,220 has CO(1--0) and
CO(3--2) luminosities of $8\times 10^9$\,K\,\kms\,pc$^2$, about twelve
times less than 14011+0252.  In the survey of 37 ultraluminous
galaxies by Solomon et al.\ (1997), the median CO luminosity was
that of Arp~220, and the dispersion was surprisingly small, only 30
per cent.  In that sample, the highest CO luminosity, for the galaxy
20087$-$0308, was only twice as high as Arp~220's, so SMM J14011+0252
definitely appears to be off-scale.

The CO(7--6) line is a new detection. The CO $J=7$ level is
$J(J+1)\times 2.77$\,K = 155\,K above the ground state, and normally
is not populated in galactic spiral-arm molecular clouds, which have
mean temperatures of 5 to 20\,K and H$_2$ densities of $\sim
200$\,cm$^{-3}$.  For example, in the central 350\,pc radius of our
Galaxy, the CO measurements by Fixsen, Bennett, \& Mather (1999) with
the {\it Cosmic Background Explorer (COBE)}, when converted to
\Lprimeco\ units, show CO(7--6) is weaker than CO(3--2) by a factor of
15. In the inner 4.8\,kpc radius ($l \pm 32^\circ$) of the Milky Way,
excluding the galactic center, the estimates by Fixsen et al.\ when
converted to \Lprimeco\ units, show CO(7--6) is weaker than CO(3--2)
by a factor of 160.  In contrast to its low level in the Milky Way,
CO(7--6) is remarkably prominent in 14011+0252, with a fifth of the
CO(3--2) luminosity. As well as the relative power, the absolute
values are very impressive: The apparent CO(3--2) luminosity is 250
times greater than that of the Milky Way, and the apparent CO(7--6)
luminosity is 8000 times greater.  The strength of CO(7--6) indicates
the CO lines in 14011+0252 are emitted in gas at $\sim 50$ to 60\,K
and more importantly, at an H$_2$ density $\sim 1700$ to
2000\,cm$^{-3}$, i.e., ten times higher than in typical spiral-arm
molecular clouds in our Galaxy.

\subsection{\it Dust Continuum at 1.3\,mm}
We detected the 1.3\,mm (224.7\,GHz) continuum, in double sideband,
and in the upper and lower sidebands separately.  All the data subsets
agree in the peak position and intensity, within the errors.  The
final 1.3\,mm continuum map (Fig.~3b) from both sidebands, excluding
CO(7--6), shows a source with a flux density of $2.5\pm 0.8$\,mJy
(1\,$\sigma$). This is about half the value of $6.1\pm 1.5$\,mJy
reported by Ivison et al.\ (2000) at 1.35\,mm.  The CO(7--6) does not
cause this discrepancy, because the line is only 225\,MHz wide, and
the bandwidth of the bolometer used by Ivison et al.\ ($>$30\,GHz)
would dilute the line flux by a factor $>$100.  We could reconcile
their results with ours if the flux-weighted centroid of their
bolometer band were at $\sim$1.15\,mm (260\,GHz) instead of 1.35\,mm.
With a source spectrum steeply rising as $\nu^{3.5}$, our narrow-band
heterodyne measurement of 2.5\,mJy at 225\,GHz would correspond to a
flux of $5\pm 2$\,mJy at 260\,GHz.

At 3\,mm (97.0\,GHz), there is no continuum, to a limit of 0.6\,mJy
(4\,$\sigma$), consistent with the 1.3\,mm emission being optically
thin radiation by dust. For a dust spectral index of 3.5, the expected
3\,mm flux density would be only 0.15\,mJy, a factor of 4 below our
limit.

\subsection{\it Position Measurement}
In the CO(3--2), (7--6), and 1.3\,mm dust maps  (Figs.\,2 and
3), the peaks are all $\sim 1.6''$ south of the initial CO(3--2)
position measured by Frayer et al.\ (1999), and $\sim 1.0''$ south
of the revised CO position listed by Ivison et al.\ (2001). For
identifying the CO with optical objects in the {\it HST} field, we
must look at both the systematic and statistical uncertainties. In
these data, the calibrators 1413+135 and 1334-127 are 11$^\circ$
and 15$^\circ$ away from 14011+0252.  With such calibrator angles,
the interferometer has an astrometric uncertainty of $\sim$0.2$''$
(see Downes et al.\ 1999, sect.4). The system noise adds
a statistical error, which for a point source is
\begin{equation}
  \Delta \theta \approx  (B/2) / (S/N)
\end{equation}
where $B$ is the beamwidth, and $S/N$ is the
signal-to-r.m.s.-noise ratio.  With natural weighting, our
CO(3--2) map has a 6$''$ beam and $S/N$ = 24, and with
uniform-weighting (Fig.~2) it has a 3.3$''$ beam and $S/N$ =
16. The best of these gives a noise error $\Delta \theta = 0.10''$
r.m.s., which convolved with the astrometric error (0.2$''$)
yields a root sum square (r.s.s.) error of 0.22$''$.
The CO(7--6) map (Fig.~3) has a 2.2$''$ beam and $S/N$ = 9.4
on the peak, so the noise error is $\Delta \theta = 0.12''$
r.m.s., which with the astrometric uncertainty (0.2$''$) gives an
r.s.s.\ error of 0.23$''$.

The new CO position thus differs significantly from those obtained by
Frayer et al.\ (1999) and Ivison et al.\ (2001), but agrees, within
the 1$\sigma$ errors, with the optical position of the J1 complex as
determined by Ivison et al.\ from the {\it HST R}-band image (Table~2
and Fig.~4).  Somewhat more puzzling is the 0.8$''$ difference between
the centroids of the CO and the 1.4\,GHz nonthermal continuum.  The
1.4\,GHz continuum is extended by 2$''$ and is likely to come from
supernovae in the starburst fueled by the molecular gas. Although
their centroids need not coincide exactly, the half-power ellipse of
the synchroton source, as determined by Ivison et al.\ (2001),
actually does include the CO peak and the optical J1 component
(Fig.~4).  The low signal-to-noise ratio (5.5$\sigma$) on the 1.4\,GHz
map by Ivison et al.\ (2001) may cause part of the apparent
discrepancy, and also explain why their new position for this VLA
source differs by 1.4$''$ from that in the Ivison et al.\ (2000)
paper.  If we increase the position errors to 2$\sigma$, then the
1.4\,GHz and CO peaks nearly overlap.  Given this uncertainty, and the
2$''$ size of the 1.4\,GHz continuum, the position difference between
the CO peak and the nonthermal peak may be less significant than it
seems at first glance.

\subsection{\it Size Measurement}
The previous report that the CO was extended over $6.6''$ (Ivison
et al.\ 2001) is not confirmed by fits to the visibilities in the
$u,v$ plane (Fig.\ 5) which give gaussian full widths to
half power of $2'' \times < 0.5''$.  The
interferometer does not ``resolve out'' more extended structure, 
because our maps
contain {\it all} of the CO flux reported by Frayer et al.\ (1999)
and Ivison et al.\ (2001). Our most compact array has
a 3\,mm beam of $7''\times 6''$, that would have responded to a
6$''$ source without missing any flux. This large-beam map 
yields the same CO flux as the small-beam map in Fig.~2. One
can also think of this in the visibility plane.  A 3\,mm source
with a size of 2.2$''$ drops to half flux at a projected baseline
of 120\,m {(Fig.~5)}.  A source three times larger would loose
half its flux at a spacing three times shorter, or 40\,m. Because
14011+0252 is at low declination, our interferometer has projected
baselines from 16\,m onward, and would have been sensitive to a
6.6$''$ source, had it existed.  We can only speculate that the CO
map by Ivison et al.\ (2001) may have too low a signal-to-noise
ratio ($S/N \sim$5.5 near the peak) to give an accurate size.  A
good rule of thumb is that one needs $S/N \geq $ 10 for a diameter
measurement.

If the size really had been $6.6''$, then even with the
magnification of 2.5 assumed by Ivison et al., the CO diameter
would have been an astonishing 20\,kpc, more than 20 times larger
than the CO-emitting regions in nearby, low-$z$ ultraluminous IR
galaxies.  It would also contradict results showing that 
objects at $z>$2 in the Hubble Deep Fields have
quite small linear dimensions 
(Ferguson, Dickinson, \& Williams 2000). These studies show that 
early-universe starbursts are efficient and small, about the size
of galaxy cores, 
not the size of complete, modern-day galactic
disks.  Even in the large, well-developed, gas-rich galaxy disks
in the present-day universe, and even in CO(1--0), the lowest
rotational transition, most of the CO is concentrated within a
central radius of 3\,kpc (see, e.g., the CO scale lengths in the
galaxy surveys by Regan et al.\ 2001, their Table~4, and
Nishiyama, Nakai, \& Kuno 2001, their Table~2).  If the CO
in 14011+0252 were really as large as 6.6$''$, then the 2$''$ beam
at 1.3\,mm would have revealed a velocity gradient of a few
hundred \kms\ across the source, but there is no gradient ---
neither in the grids of spectra (Fig.\,6) nor in the maps of
individual CO velocity channels (Fig.\,7).

Although the CO region is not as large as originally claimed, its
measured diameter of 2$''$ (17\,kpc) is still unlikely to be a true
size. Merely detecting CO(7--6) makes such a large CO extent very
improbable.  Because the CO J=7 level is 155\,K above the ground
state, seeing the line at all implies a gas kinetic temperature of at
least 50\,K and an average H$_2$ density $>1500$\,cm$^{-3}$.  These
temperatures and mean densities are found in the circumnuclear
starburst disks of radius $\sim 300$\,pc in the centers of ULIGs, but
they are implausible over 17\,kpc.

One could imagine a giant, spread-out, opaque ``reservoir'' of gas and
dust hiding several simultaneous ULIG-type starbursts, but the data 
suggest most of the CO flux comes from a single object: (1) a
single gaussian can fit the visibility amplitudes (Fig.\ 5); (2) the
CO channel maps (Fig.\,7) have only a single component, at the same
position, within the noise, at all velocities; (3) the grids of the
spectra (Fig.\,6) have no obvious signs of multiple velocity features.
In fact, both the linewidth (190\,\kms ) and simple profile shape of
the CO spectra (Fig.\,1) resemble those in Mrk~231 (230\,\kms ), which
has a single, circumnuclear disk of radius 460\,pc (Bryant \& Scoville
1996; Downes \& Solomon 1998).  The CO(2--1) brightness temperature in
Mrk~231 is 29\,K in a 0.7$''$ beam, corresponding to a true brightness
temperature of $\sim 50$\,K.

\section{IS SMM J14011+0252 STRONGLY LENSED?}
We suspect 14011+0252 is lensed not only by the core of the
foreground cluster of galaxies, but also by a galaxy along the
line of sight.  We estimate the magnifying factor directly from
the surface brightness. Since the CO is optically thick, its
brightness temperature must be comparable to the gas kinetic
temperature. The CO(7--6) detection tells us the gas is warm and
dense, and its strength relative to CO(3--2) indicates the gas
temperature.  This {\it intrinsic} brightness temperature, divided
by $1+z$, is what one would expect if the source filled the beam.
The ratio of {\it observed} to expected brightness temperatures is
then the area filling factor in the beam. Gravitational lensing
preserves brightness temperature, so from the measured CO major
axis, one obtains the minor axis, and their ratio is roughly the
magnifying factor.  Another way to derive this is to start from
the apparent CO luminosity, which is
\begin{equation}
\LprimeCO ({\rm obs}) \ \ = \ \  363\,
(S\Delta V)\, \lambda^2\,  D^2_A\, (1+z)\ \ \ ,
\end{equation}
where $(S\Delta V)$ is the integrated CO flux in Jy\,\kms ,
$\lambda$ is the redshifted wavelength in mm, and $D_A$ is the
angular diameter distance in Mpc (this is a variant of eq.\,[3] of
Solomon, Downes, \& Radford 1992).  If the true geometry is a
gaussian with half-power diameter $d$, then the magnification of
the CO is
\begin{equation}
m_{\rm CO} \ \ = \ \ {d_m\over d}\ \ = \ \ {{1.133\,
d_m^{\,2}\,\Delta V}\over \LprimeCO ({\rm obs})} \, f_V \,T_b \ \
\ \ ,
\end{equation}
where $d$ is the true CO diameter in pc, $d_m$ is the magnified
major axis in pc, $\Delta V$ is the linewidth in \kms ,
$\LprimeCO$ is the amplified CO(3--2) luminosity in \Kkmspc ,
$f_V$ is the velocity filling factor, and $T_b$ is the CO(3--2)
rest frame brightness temperature, which the CO(3--2)/CO(7--6)
ratio indicates is $35\pm 5$\,K (this is a version of eq.\,[2] of
Downes, Solomon, \& Radford 1995). For 14011+0252, the data imply
that {\it if the measured CO major axis is $2''$, then the
gravitational lens magnifies the CO size by a factor of 25}. The
flux amplification inferred from our brightness temperature
argument is for the sum of all the CO emission, which could be in
a single, long arc or in several multiple-image spots or arclets.

Here is another way to present this argument:

1) In many respects, the spectral energy distribution of
14011+0252 resembles closely that of Arp 220 and other
local-universe ULIGs, with much of the input power being
re-radiated by dust at 40 to 65\,K. Because its molecular gas also
appears to be heated by an extraordinary starburst, 14011+0252
should have an intrinsic CO brightness temperature comparable with
Arp 220's.

2) On interferometer maps of Arp 220 (Scoville, Yun \& Bryant 1997;
Downes \& Solomon 1998), one directly measures average CO(1--0) and
(2--1) brightness temperatures of 35\,K, so this is a reasonable value
for the intrinsic CO brightness temperature in 14011+0252.

3) We may also {\it calculate} a mean CO brightness temperature of Arp
220 from its total CO luminosity, linewidth, and size, for a velocity
filling factor $f_V$=1, i.e., a mean $T_b$ averaged over the line
profile.  This mean brightness temperature is also 35\,K, in agreement
with the direct measurement off the Arp 220 CO maps.  Again, this is
the expected intrinsic mean brightness temperature for the CO in
14011+0252.

4) In 14011+0252, the \Lprimeco\ (3--2)/(7--6) ratio is $5\pm 1$,
corresponding to a molecular gas kinetic temperature of $\sim 50$
to 60\,K, and an H$_2$ density of $\sim 1700$ to 2000\,cm$^{-3}$.
In an escape probability model, these parameters yield a predicted
CO(3--2) brightness temperature of $35\pm 5$\,K, confirming the
brightness temperature deduced by analogy with Arp 220 and other
ultraluminous galaxies.

5) Because gravitational lensing conserves surface brightness, the
mean CO(3--2) brightness temperature should be 35\,K, whatever the
magnified size.

6) The measured major axis is $2.2''\pm 0.4''$ (gaussian FWHP), and
from the intrinsic brightness temperature, linewidth, and {\it
apparent} CO(3--2) luminosity, the derived minor axis turns out to be
$0.083''\pm 0.015''$.  The major/minor axis ratio is thus $25\pm 5$,
and this is how much the lens magnifies the CO size.

Here's a shortcut to this answer:

1) From the observed CO(3--2) luminosity, linewidth, and size
of 2.2$''$, the brightness temperature would be only 0.37\,K, if the
emitting area were circular. Correcting by $(1+z)$, we would get
1.3\,K for the apparent $T_b$ at $z$=2.6.

2) This apparent $T_b$ of 1.3\,K is $25\pm 5$ times smaller than
the true $T_b$ of $35\pm 5$\,K expected from the CO line ratio and
the similarity of the far-IR spectrum to Arp 220's. Again, this
factor of 25 is how much the lens amplifies the CO flux.

Table~2 lists the CO and dust results for 14011+0252. We also give the
true radius, gas mass, dynamical mass, and far-IR luminosity after
correcting downward by a factor of 25, which we think is the combined
lensing effect of an individual galaxy and the cluster ensemble.  The
magnifying factor would of course be lower if several different,
simultaneous, ULIG-level starburst sources gave area and velocity
filling factors that diluted the brightness temperature in the beam.
For the reasons given in sect.\ 2.5 --- simple gaussian visibility
function, one sole component in channel maps, one sole gaussian
spectral profile --- we prefer to treat the CO flux as coming from a
single object.

\section{A NEW INTERPRETATION}
The CO brightness argument for strong lensing now suggests a
different origin for some of the IR/optical features found by
Ivison et al.\ (2001).  Rather than being interacting galaxies and
tidal tails of a merger, some of these bright knots may instead be
multiple images. If for example, the J1(southeast) and J2
components are really the same object, this would explain:

--- why J1 and J2 have identical redshifts in Ly$\alpha$, and why they
also have the same UV absorption lines;

--- why J1 and J2 {\it both} have Ly$\alpha$ shifted by $-$400\,\kms\
from the lower-excitation H$\alpha$ and the CO.  Ivison et al.\ (2000;
2001) interpret J1 and J2 as interacting galaxies, and the 400\,\kms\
shift as an outflow. But Ly$\alpha$ is at the {\it same} velocity in
both J1 and J2. Why would two interacting galaxies both have a
blueshifted outflow? And why would the two outflow velocity vectors be
at the same angle to the line of sight so as to yield a shift of
400\,\kms\ in both galaxies?

--- why J2 looks like an arc in the high-resolution {\it HST} picture
 (Ivison et al.\ 2001, their Fig.~1, middle).

--- why the colors of J1 and J2 are different on the low-resolution
UKIRT+{\it HST} composite picture (Ivison et al.\ 2001, their Fig.~1, 
left): --- the simplest answer is that the main blob in the J1 complex
is the lensing galaxy, and it is redder than 14011+0252.

Different intrinsic {\it areas} at the emitted wavelengths may
explain part of the different colors of J1, J2, and J1n.  The observed
{\it K}-band is rest-frame 6000\,\AA , including starlight from the
nuclear bulge and H$\alpha$+[N II] from H~II regions.  The observed
{\it U}-band is rest-frame 1000\,\AA , which comes from the more
extincted, and possibly more confined, hot-star continuum and
Ly$\alpha$ outflow. These different-sized regions would have different
magnifications in the main and secondary images. Color differences
among main and counter-images, or between arcs and bright spots, are
known in other objects, such as MG 0414+0534 (Lawrence et al.\ 1995;
Falco, Leh\'ar, \& Shapiro 1997).

In fact, the {\it HST} image of 14011+0252 in Fig.~8, left (from
Ivison et al.\ 2001, their Fig.~1, middle) resembles that of the quasar
MG 0414+0534 in Fig.~8, right (from Falco et al.), which has four
bright spots in the ``classic'' lensed configuration of a bright,
close double A1+A2 separated by $0.4''$, plus two fainter spots B and
C, each $2''$ from A1+A2 and from each other (e.g., Falco et al.;
Blandford \& Narayan 1992, their Fig.~6c,d).  As in 0414+0534, the
central blob in the J1 component of 14011+0252 may actually be the
lensing galaxy, and it is this galaxy that may cause the color
difference between J1 and J2 on the smoothed ``true color" picture by
Ivison et al.\ (2001; their Fig.~1, left). Besides the possibly analogous
quadruple spots or arclets in the optical, there are other points in
common.  Like 14011+0252, the galaxy 0414+0535 is at redshift 2.6, has
H$\alpha$ and H$\beta$, is very dusty and red, with $r-K$ = 7\,mag
(e.g., Hewitt et al.\ 1992; Lawrence et al.\ 1995), and has a CO(3--2)
power (Barvainis et al.\ 1998) identical to that of 14011+0252.  The
source 0414+0534 has a quasar, and we think its strong CO line comes
from the same kind of circumnuclear, molecular-gas disk as in
local-universe ULIGs containing quasars, like Mrk~231.

The CO in 14011+0252 may also be like the arcs \#384/468 lensed by
the $z=0.175$ galaxy cluster A2218 (Pell\'o et al.\ 1988; 1992;
Kneib et al.\ 1996).  The remarkable arc \#384 stretches over
4$''$ and appears to be two ionized regions distorted into long
parallel tracks, each about 0.25$''$ wide.  The arc is at $z$ =
2.515, and its $R$-band flux is amplified by $2.9\pm 0.3$ mag
(Ebbels et al.\ 1996), or a factor of $14.5\pm 2$.  That is, the
length, redshift, and magnification of the arc \#384 in A2218 all
resemble those of the CO in 14011+0252.

Figure~9 shows a type of model that may explain 14011+0252.  In the
source plane, the radial and tangential caustics are not centered on
the lensing galaxy, but are pulled off-center by the mass of the
$z=0.25$ cluster Abell 1835, whose dominant cD galaxy is 50$''$ to the
northwest.  In the model, the galaxy 14011+0252 at $z=2.6$ has a
compact UV core with a diameter of 0.02$''$ (black core in Fig.~9,
left), surrounded by a larger molecular ring (grey halo in the
Figure). The core UV light is deflected into two small arcs (the black
central parts of the arcs in Fig.~9, right), suggestive of the J1-SE
and J2 components in the {\it HST R}-band field of 14011+0252.  In
lens plane, the millimeter CO and dust radiation is elongated into the
larger grey parts of the arcs in Fig.~9.  Most of the CO and dust
signal is in the larger eastern arc, which we associate with J1.  For
the figure, we did not try to find the best model, but simply adopted
a velocity width of 200\,\kms\ and ellipticity of 0.2 for the core of
the lensing galaxy to illustrate what may be happening.  More
sensitive, higher-resolution CO maps are needed to derive the relative
deflecting strengths of the main lensing galaxy and the cluster as a
whole.

Table~3 lists the observed and de-magnified CO parameters of
14011+0252 and 0414+0534.  For 0414+0535, we used the CO(3--2)
data of Barvainis et al.  We also give the values for Arp\,220, if
moved to $z=2.6$, based on the CO(3--2) intensity (Mauersberger et
al.\ 1999) and interferometer size measurements (Downes \& Solomon
1998).  The table shows that with our model of comparable CO
brightness temperatures and hence strong lensing, the intrinsic
properties of the molecular disks in 14011+0252 and 0414+0535 must
be roughly similar to those in Arp\,220 and other ultraluminous IR
galaxies in the local universe.

Strong lensing may also explain another curiosity about this
galaxy. Adelberger \& Steidel (2000) showed 14011+0252 obeys
nicely the rest-frame UV, sub-mm, and cm-radio correlations
established for rapidly star-forming galaxies at low redshifts.
None of the $\sim$800 spectroscopically confirmed $z\sim 3$
galaxies in their UV-selected sample, however, is as bright in the
near-IR as 14011+0252 ($R$-magnitude 21.25) and $\sim 95$\% of the
galaxies in their sample are ten times fainter, implying dust
fluxes ten times lower than SMMJ14011's, or $\leq 1$\,mJy at
850\,$\mu$m. Most of their UV-selected galaxies would thus be
undetectable at 1.3\,mm in either the dust continuum or in CO. Our
interpretation that the lens magnifies the source by more than an
order of magnitude means that 14011+0252 is basically the same
type of dusty, rapidly star-forming galaxy as many of the other
UV-selected galaxies at high redshift.

\section{CONCLUSIONS}
Contrary to an apparent result that the CO in 14011+0252 is
extended by 6.6$''$, suggesting a starburst on a much larger scale
than in local ultraluminous IR galaxies, new interferometer
measurents yield a size of only $2''\times \leq 0.5''$ for the CO
and dust emission.  Our brightness-temperature argument further
indicates the CO size must be magnified by a factor of 25.  The 
gravitational lens may be an isolated galaxy along the line of
sight, whose effects add to those of the foreground cluster A1835.

The image configuration in 14011+0252 resembles that in the
gravitationally lensed quasar 0414+0535, and the circumnuclear
starburst disks in both galaxies are probably magnified by comparable
amounts.  This means the true radii, gas masses, star formation rates,
and far-IR luminosities of their molecular disks are all about the
same as those in local-universe ultraluminous IR galaxies like
Arp\,220.  They are not huge, very massive disks in the early universe.

\begin{acknowledgements}
We thank the operators on Plateau de Bure for their help in observing,
R.\ Lucas for his aid with the visibility plots, M.\ Bremer for help
with the figures, R.\ Neri for the lens model, and the referee for
very useful comments.
\end{acknowledgements}

\newpage
\begin{deluxetable}{lccc}
\tablecaption{POSITION MEASUREMENTS}
\tablehead{
        &\colhead{R.A.}     &\colhead{Decl.}
                &\colhead{Refs. \tablenotemark{a}}
\\
\colhead{Source}    &\colhead{(J2000)}  &\colhead{(J2000)}  &
}
\startdata
\multicolumn{4}{l}{ SMM J14011+0252: }
\\
\multicolumn{4}{l}{ {\it Millimeter data:} }
\\
CO(3--2)
    &14$^{\rm h}$01$^{\rm m}$04.93$^{\rm s} \pm 0.02^{\rm s}$
    &02$^\circ52'24.1'' \pm 0.2''$
    &1
\\
CO(7--6)
    &14$^{\rm h}$01$^{\rm m}$04.92$^{\rm s} \pm 0.02^{\rm s}$
    &02$^\circ52'23.8'' \pm 0.3''$
    &1
\\
1.3\,mm dust
    &14$^{\rm h}$01$^{\rm m}$04.93$^{\rm s} \pm 0.03^{\rm s}$
    &02$^\circ52'24.5'' \pm 0.5''$
    &1
\\
\\
\multicolumn{4}{l}{ {\it IR and centimeter data:} }
\\
$I$-band, J1 component
    &14$^{\rm h}$01$^{\rm m}$04.95$^{\rm s} \pm 0.02^{\rm s}$
    &02$^\circ52'24.0'' \pm 0.3''$
    &2
\\
1.4\,GHz VLA peak
    &14$^{\rm h}$01$^{\rm m}$04.92$^{\rm s} \pm 0.01^{\rm s}$
    &02$^\circ52'24.8'' \pm 0.2''$
    &2
\\
\\
\multicolumn{3}{l}{Phase calibrators:}
\\
1413$+$135
        &14$^{\rm h}$15$^{\rm m}$58.81749$^{\rm s} \pm 0.00009^{\rm s}$
    &$+13^\circ 20'23.713'' \pm 0.003''$    &3
\\
1334$-$127
    &13$^{\rm h}$37$^{\rm m}$39.78278$^{\rm s} \pm 0.00005^{\rm s}$
    &$-12^\circ 57'24.693'' \pm 0.001''$    &3
\\
\\
\enddata
\tablenotetext{a}{
{\it References}: (1) This paper;
(2)\ Ivison et al.\ (2001); (3) Ma et al.\ (1998).}
\\
\end{deluxetable}
\clearpage
%
\newpage
\begin{deluxetable}{lccl}
\tablecaption{CO LINE AND DUST PROPERTIES OF SMM J14011+0252}
\tablehead{
\colhead{Parameter}&\colhead{CO(3--2)}&\colhead{CO(7--6)}
}
\startdata
\multicolumn{3}{l}{{\it Observed CO quantities:}} &{\it Unit}
\\
Center frequency    &96.989         &226.251&GHz
\\
Redshift (lsr)          &$2.5652\pm 0.0001$     &$2.5651\pm 0.0002$ &---
\\
CO peak flux \tablenotemark{a}
            &$13.2\pm 1$        &$12.4\pm 3$    &mJy
\\
Linewidth           &$190\pm 11$        &$170\pm 30$    &\kms
\\
CO integrated flux      &$2.8\pm 0.3$       &$3.2\pm 0.5$   &Jy\,\kms
\\
Apparent
$L^\prime_{\rm CO}$     &$10.8\pm 1$    &$2.3\pm 0.3$   &10$^{10}$\,\Kkmspc
\\
$L^\prime$ ratio (3--2)/(7--6)
            &$4.8\pm 1$     &---        &---
\\
Major axis FWHP     &$2.2\pm 0.4$       &$2.3\pm 0.5$   &arcsec
\\
Minor axis FWHP     &$<0.5$             &$<0.8$     &arcsec
\\
Position angle      &$10\pm 20$         &$0\pm 30$  &deg
\\
\\

\multicolumn{3}{l}{{\it Derived CO quantities:} \tablenotemark{b}}
\\
Intrinsic CO $T_b$  &$35\pm 5$      &$7\pm 1$   &K
\\
Lens magnification  &$25\pm 5$      &$23\pm 5$  &---
\\
True $L^\prime_{\rm CO}$&$4.4\pm 0.9$       &$1.0\pm 0.3$ &$10^9$\,\Kkmspc
\\
Gas mass $M$(H$_2$+He) \tablenotemark{c}
            &$3.5\pm 0.7$       &---         &$10^9$\,\Msun
\\
True radius, $R$     &$360\pm 100$         &$420\pm 130$   &pc
\\
Dynamical mass $RV^2_{\rm rot}/G$ \tablenotemark{d}
            &$6\pm 2$   &---&$10^9$\,\Msun
\\
\\
\multicolumn{3}{l}{{\it Dust quantities:}}
\\
Dust flux       &$<0.6$ (4\,$\sigma$)&$2.5\pm 0.8$&mJy
\\
Dust mass       &---    &0.6---2.4          &$10^7$\,\Msun
\\
True $L_{\rm FIR}$  &---    &$9\pm 3$       &$10^{11}$\,\Lsun
\enddata
\tablenotetext{a}{\,CO peak fluxes are for beams of $7.1''\times 5.6''$
at CO(3--2) and $2.2''\times 2.0''$ at CO(7--6).}
\\
\tablenotetext{b}{\,Adopted luminosity distance = 22.6\,Gpc ($H_0
=$65\,\kms\,Mpc$^{-1}$, $\Omega_m =$ 0.3, $\Omega_\Lambda =$
0.7); angular diameter distance = 1.781\,Gpc; linear
scale: 1$'' \leftrightarrow 8635$\,pc.}
\tablenotetext{c}{
\,Gas mass = $\pi R^2 n(H_2) 1.36
m(H_2)\Delta z$, where we took a mean H$_2$ number density
of 2000\,cm$^{-3}$ in a disk of true radius $R$ and height $\Delta z$ of
50\,pc.  This gives a ratio $M_{\rm gas}/ L^\prime_{\rm CO}$ $\approx
0.8$\,\Msun\,(K \kms \,pc$^2$)$^{-1}$, as deduced for the centers of ULIGs
(Downes \& Solomon 1998).
}
\tablenotetext{d}{\,For the dynamical mass within radius $R$,
we assumed a disk inclined at 50$^\circ$, rotating at 250\,\kms .}
\end{deluxetable}
\clearpage
%
\newpage
\begin{deluxetable}{lcccl}
\tablecaption{SOURCE COMPARISON, FOR $z=2.6$}
\tablehead{
\colhead{Source: \tablenotemark{a} }
&\colhead{SMM J14011} &\colhead{MG 0414}    &\colhead{Arp\,220 moved}\\
&\colhead{+0252}      &\colhead{+0534}          &\colhead{to $z$=2.6}}
\startdata
{\it Observed Line:}
        &CO(3--2)   &CO(3--2)   &CO(3--2)   &{\it Unit} \\
Redshift (lsr)  &2.5653     &2.639  &2.6        &---        \\
Peak flux density
        &13         &4.4    &0.2        &mJy        \\
Linewidth   &190        &580    &480        &\kms       \\
Integrated flux &2.7        &2.6        &0.2        &Jy\,\kms   \\
Major axis  &2.2        &(1.1)      &0.09       &arcsec     \\
Minor axis  &$<0.5$     &$<1.1$     &0.03       &arcsec     \\
Apparent $L^\prime_{\rm CO}$
        &108        &100    &8  &10$^9$\,\Kkmspc    \\  \\
\multicolumn{4}{l}{{\it Derived CO quantities:}}            \\
CO magnification &25        &(20)     &1        &---        \\
True radius, $R$ &360        &(235)      &400       &pc     \\
True $L^\prime_{\rm CO}$
                &4.4        &5.0        &8  &10$^9$\,\Kkmspc    \\
Dynamical mass  $M_{\rm dyn}$ \tablenotemark{b}
                &6          &5.9        &10 &10$^9$\,\Msun      \\
Gas mass $M_{\rm gas}$ \tablenotemark{c}
        &3.5        &1.1        &5.0    &10$^9$\,\Msun      \\
$M_{\rm gas}$ /$L^\prime_{\rm CO}$
        &0.8        &0.2    &0.6        &---        \\ \\
\multicolumn{4}{l}{{\it Dust quantities:}}              \\
1.3\,mm flux density
        &2.5    &20.7 nt        &0.4        &mJy    \\
Apparent $L_{\rm FIR}$
        &23    &---     &1.5        &10$^{12}$\,\Lsun \\
True $L_{\rm FIR}$
        &1.7       &---     &1.5        &10$^{12}$\,\Lsun
\\
\enddata
\tablenotetext{a}{\,For SMM J14011, parameter uncertainties are
listed  in Table 2. For MG 0414, the data are from Barvainis et al.\ 1998,
and the adopted magnification from Barvainis \& Ivison 2002.
The 1.3\,mm continuum of MG 0414 is nonthermal (nt), not dust.}
\\
\tablenotetext{b}{\,For $M_{\rm dyn}(<R)$, the adopted rotation
velocities (\kms) and inclinations are (250\, 50$^\circ$), (330,
80$^\circ$), and (330, 40$^\circ$).}

\tablenotetext{c}{\,$M_{\rm gas}$ is for H$_2$+He, a disk height =50\,pc,
 and $n$(H$_2$) =2000\,cm$^{-3}$.}
\end{deluxetable}
\clearpage
\begin{figure}
\vspace{-1.5cm}
\includegraphics{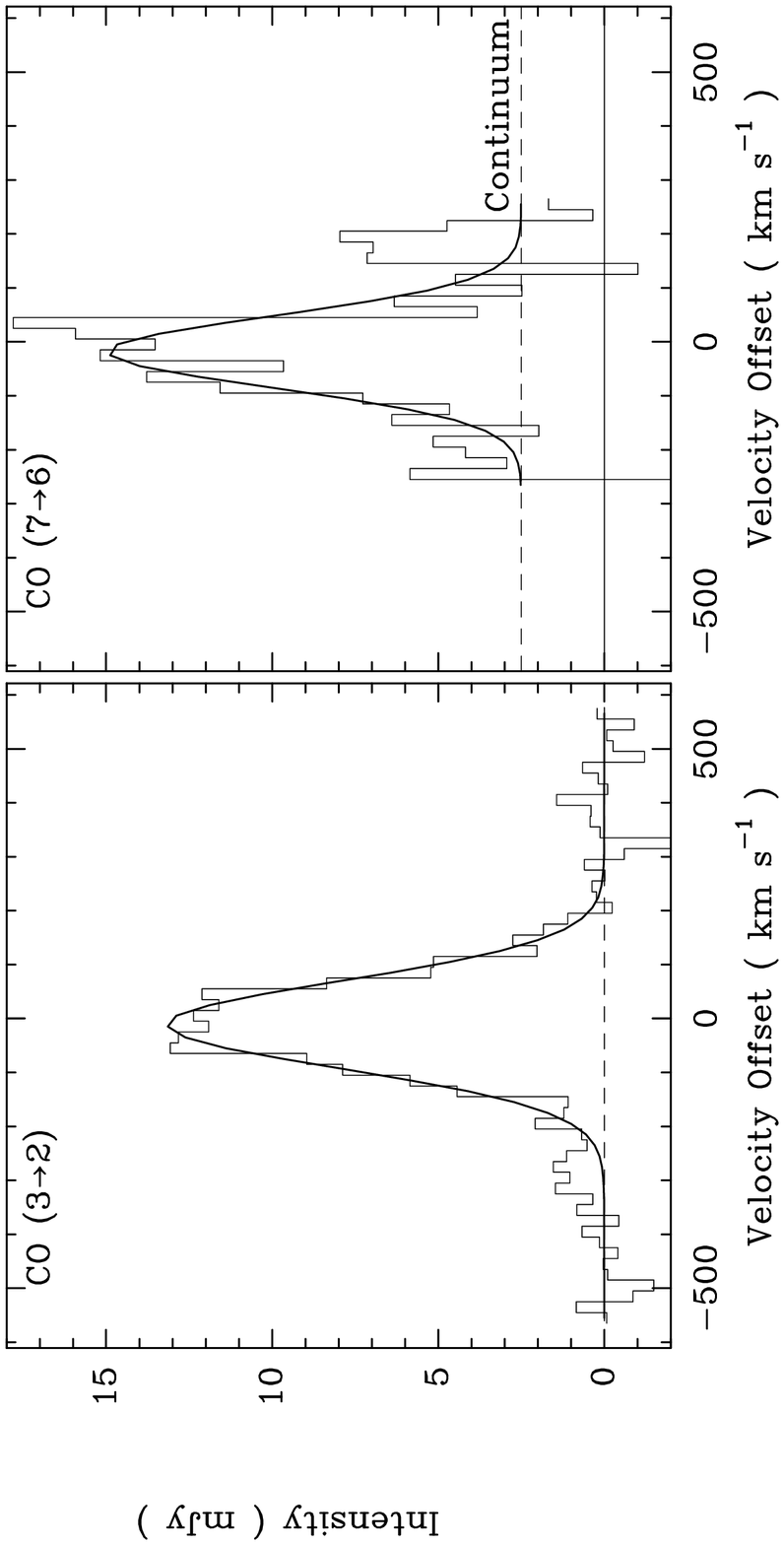}
\end{figure}
\begin{figure}
\vspace{2.5cm}
\caption[CO(3--2) and (7--6) spectra at peak.]
{CO spectra with 20\,\kms\ resolution, at the peak position
(Table~1) of SMM J14011+0252, with their gaussian fit curves
superposed.

{\it  Left:} CO(3--2) spectrum. The r.m.s.\ noise is 0.84\,mJy,
and the beam is $7.1'' \times 5.6''$ with $T_b/S$ = 3.3\,K/Jy. The
line peak is 13.2\,mJy, or 0.043\,K. Velocity offsets are relative
to 96.989\,GHz.

{\it  Right:}
CO(7--6) spectrum.  The r.m.s.\ noise is 3.1\,mJy, and the beam is
$2.2'' \times 2.0''$ with $T_b/S$ = 5.4\,K/Jy.  Velocity offsets are
relative to 226.251\,GHz. The line is above a 2.5\,mJy continuum
(measured in the other sideband).}
\end{figure}
\clearpage
%
\newpage
\begin{figure}
\plotone{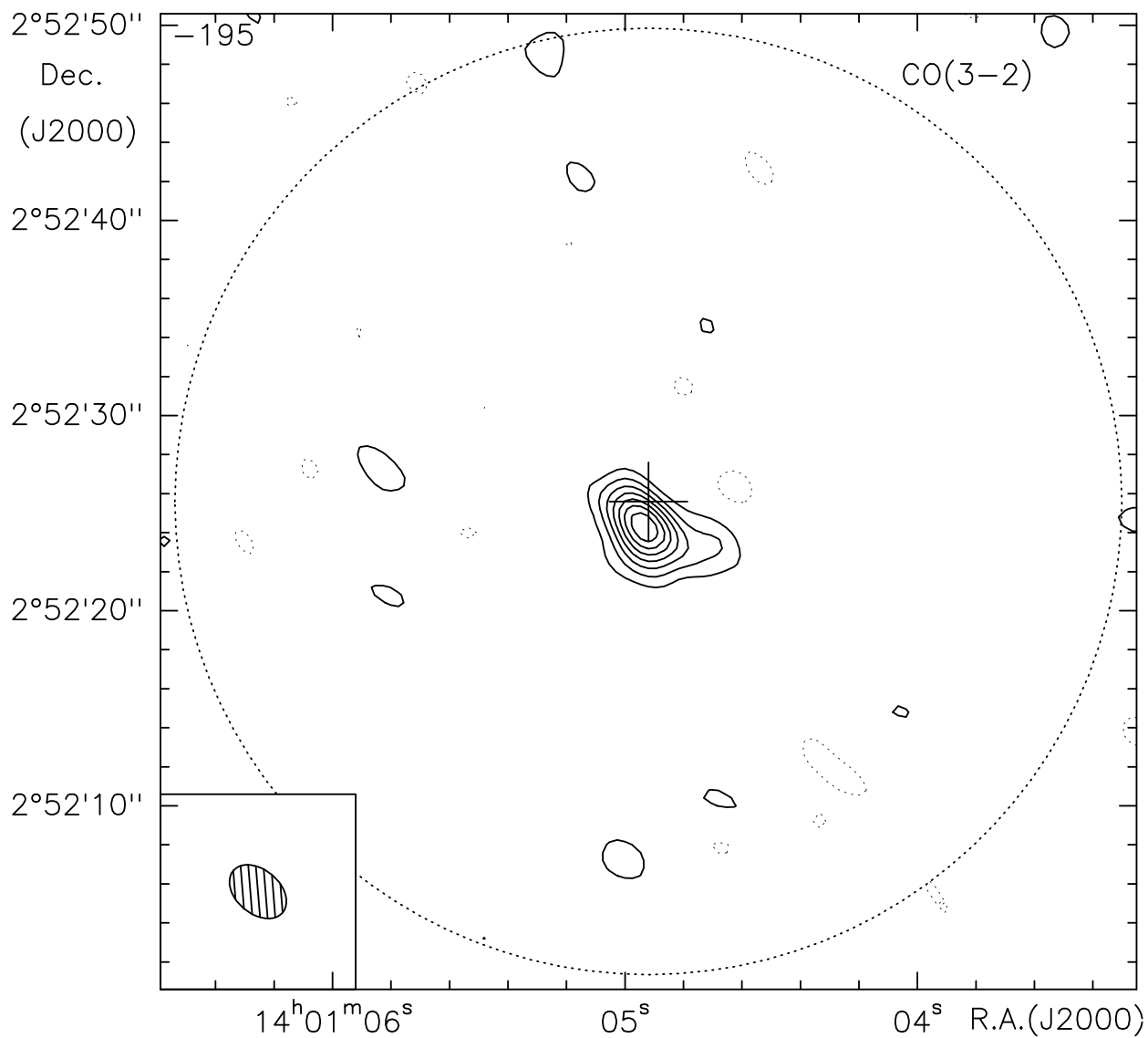}
\caption[3mm CO(3--2) map]
{Map of CO(3--2) integrated
over 260\,\kms , from SMM J1401+0252.
The beam is $3.3'' \times 2.3''$ at
p.a.\ 50$^\circ$ (lower left), with $T_b/S$ = 17.2\,K/Jy.  The large dotted
circle shows the 49$''$ primary lobe. The contour step is
 0.82\,mJy\,beam$^{-1}$ ($2\sigma$), or
 0.21\,Jy\, beam$^{-1}$ \kms\ in velocity-integrated flux;
negative contours are dashed, zero level omitted. The cross marks
 the initial CO position (Frayer et al.\ 1999), that we adopted as
the phase center for our observing.}
\end{figure}
%
\begin{figure}
\plottwo{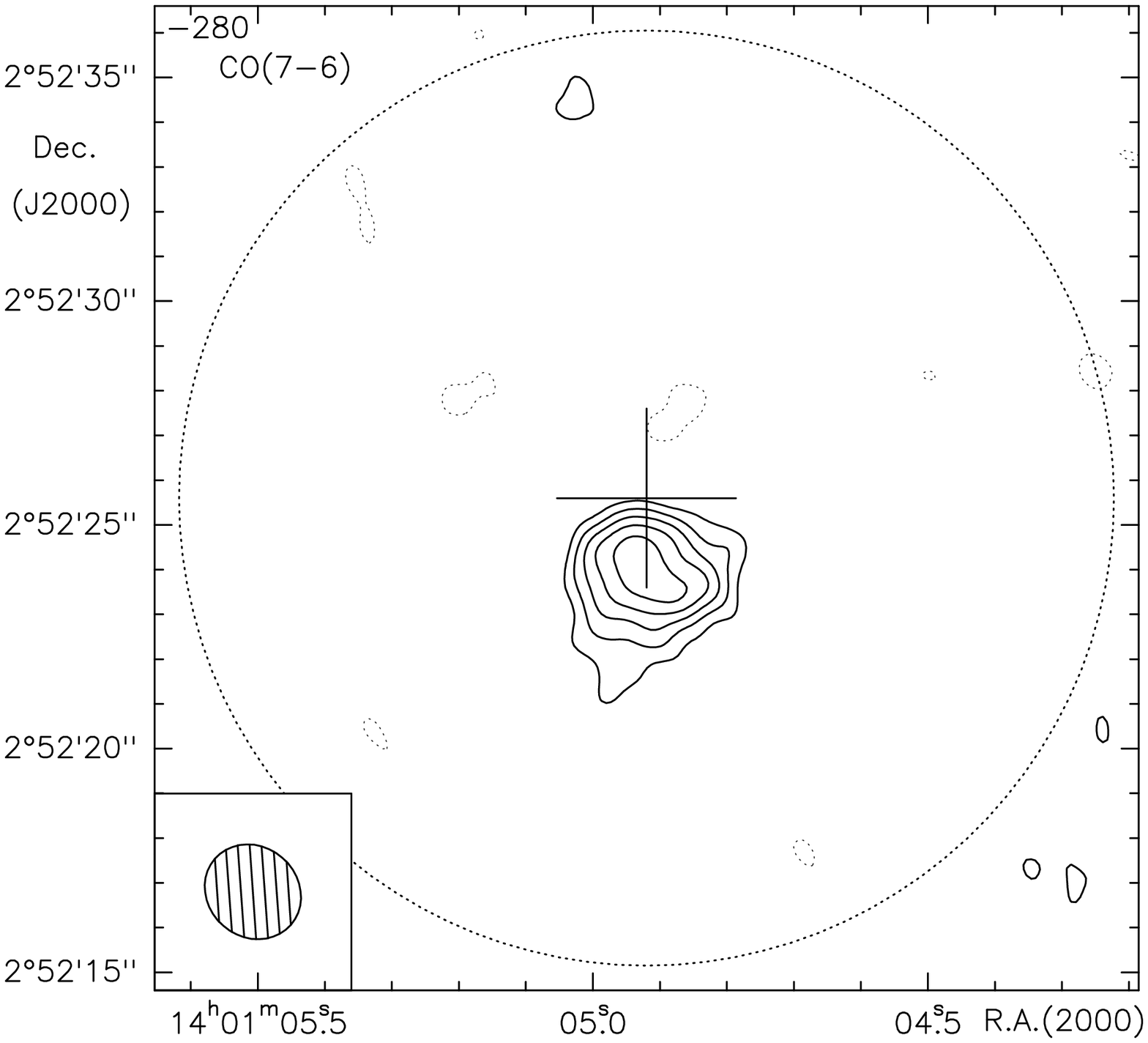}{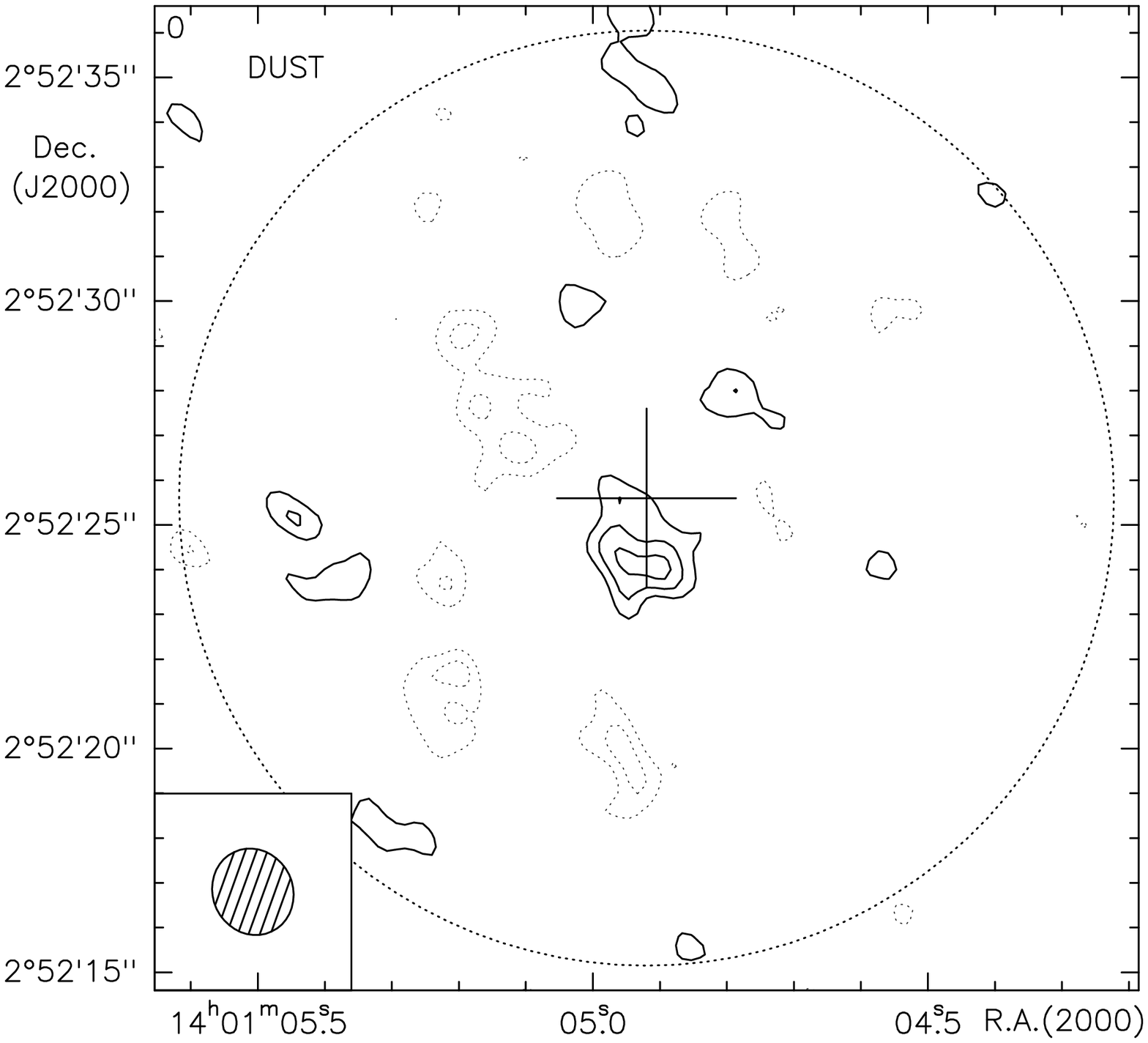}
\caption[1mm CO(7-6) and dust map] {Maps of CO and dust at
1.3\,mm.

{\it  Left:} CO(7--6) integrated over the central 160\,\kms\ of
the profile. Contours start at 2.6\,mJy beam$^{-1}$ ($2\sigma$)
and increase in steps of 1.3\,mJy beam$^{-1}$ ($1\sigma$), or a
step in velocity-integrated flux of 0.21\,Jy\, beam$^{-1}$ \kms ;
negative contours are dashed, zero level omitted.  On this map,
the continuum peak of 1.6\,mJy would be below the first contour.
The beam is 2.2$'' \times 2.0''$ at
p.a.\ 48$^\circ$ (lower left), with $T_b/S$ = 5.4\,K/Jy.

{\it  Right:} Dust continuum at 1.3\,mm (224.7\,GHz). Contours
start at 0.8\,mJy beam$^{-1}$ ($2\,\sigma$) and increase in steps
of 0.4\,mJy ($1\,\sigma$); negative contours are dashed, zero
level omitted. The spatially-integrated flux density is 2.5\,mJy.
The beam is 2.0$'' \times 1.8''$ at p.a.\ 25$^\circ$ (lower left),
with $T_b/S$ = 6.7\,K/Jy. In each diagram, the large dotted circle
shows the 20.9$''$ primary lobe, and the cross marks the initial
CO position from Frayer et al.\ (1999). }\end{figure} \clearpage
%
\begin{figure}
\caption[CO (3--2) and (7--6) peaks compared with VLA peak] {CO
contours superposed on the $HST$ F702W image from Ivison et al.\
(2001), and compared with VLA halfwidth at 1.4\,GHz.

{\it Left:} CO(3--2) map, with same contours as in Fig.~2, beam 
$3.3''\times 2.3''$ at p.a.\ 50$^\circ$.

{\it Right:} CO(7--6) map, with same contours as in Fig.~3a, 
beam $2.2\times 2.0''$ at p.a.48$^\circ$.

In each diagram, the dark ellipse shows the 2.3$''$ $\times 1.5''$
half-power widths of the 1.4\,GHz non-thermal VLA source, as derived
by Ivison et al.\ (2001).  The 1-$\sigma$ error quoted by those
authors for the optical position is $\pm 0.3''$ (small circle above
the F702W label).

}\end{figure}
\clearpage
\newpage
%
\begin{figure}
\vspace{1.5cm}
\plottwo{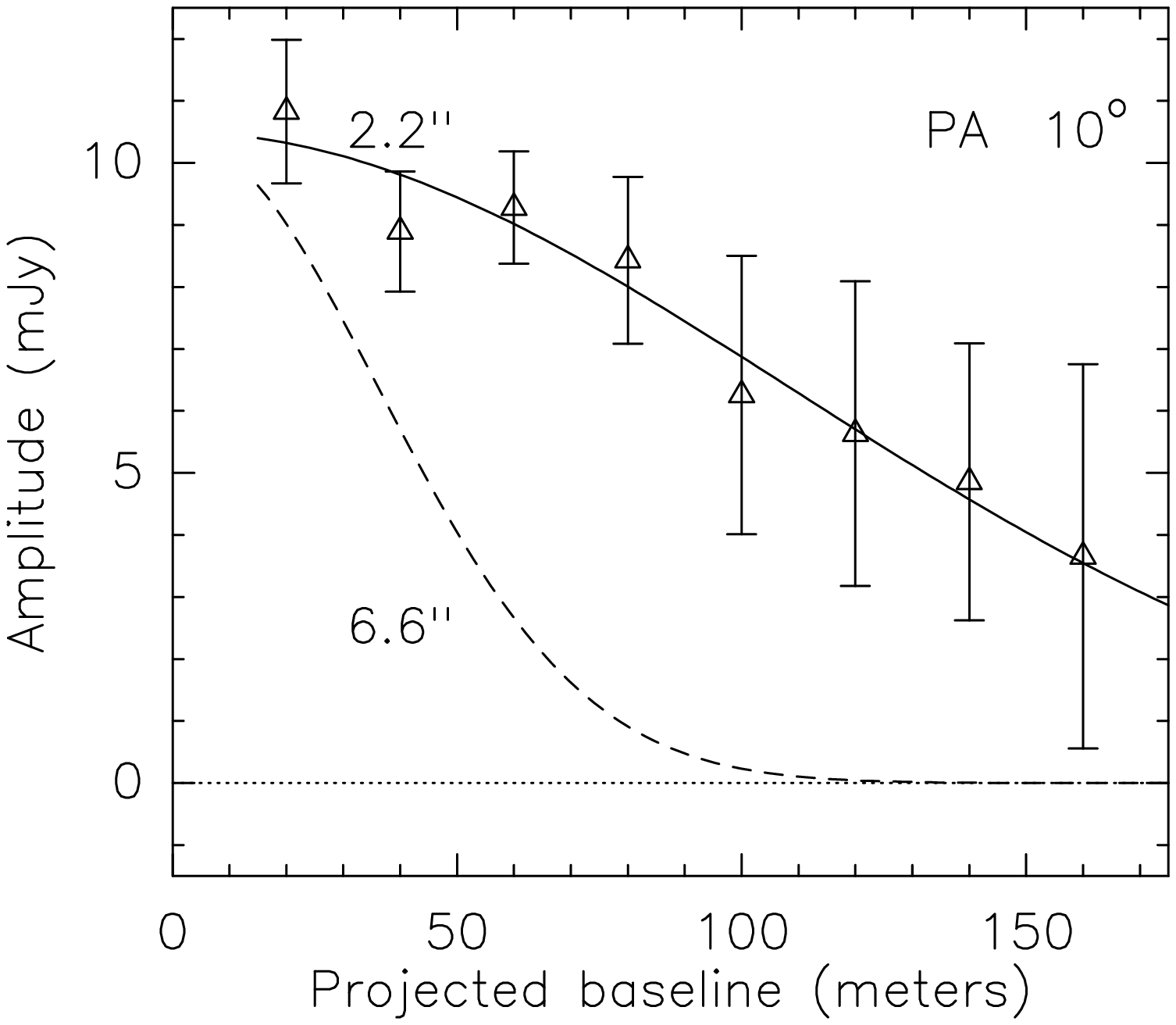}{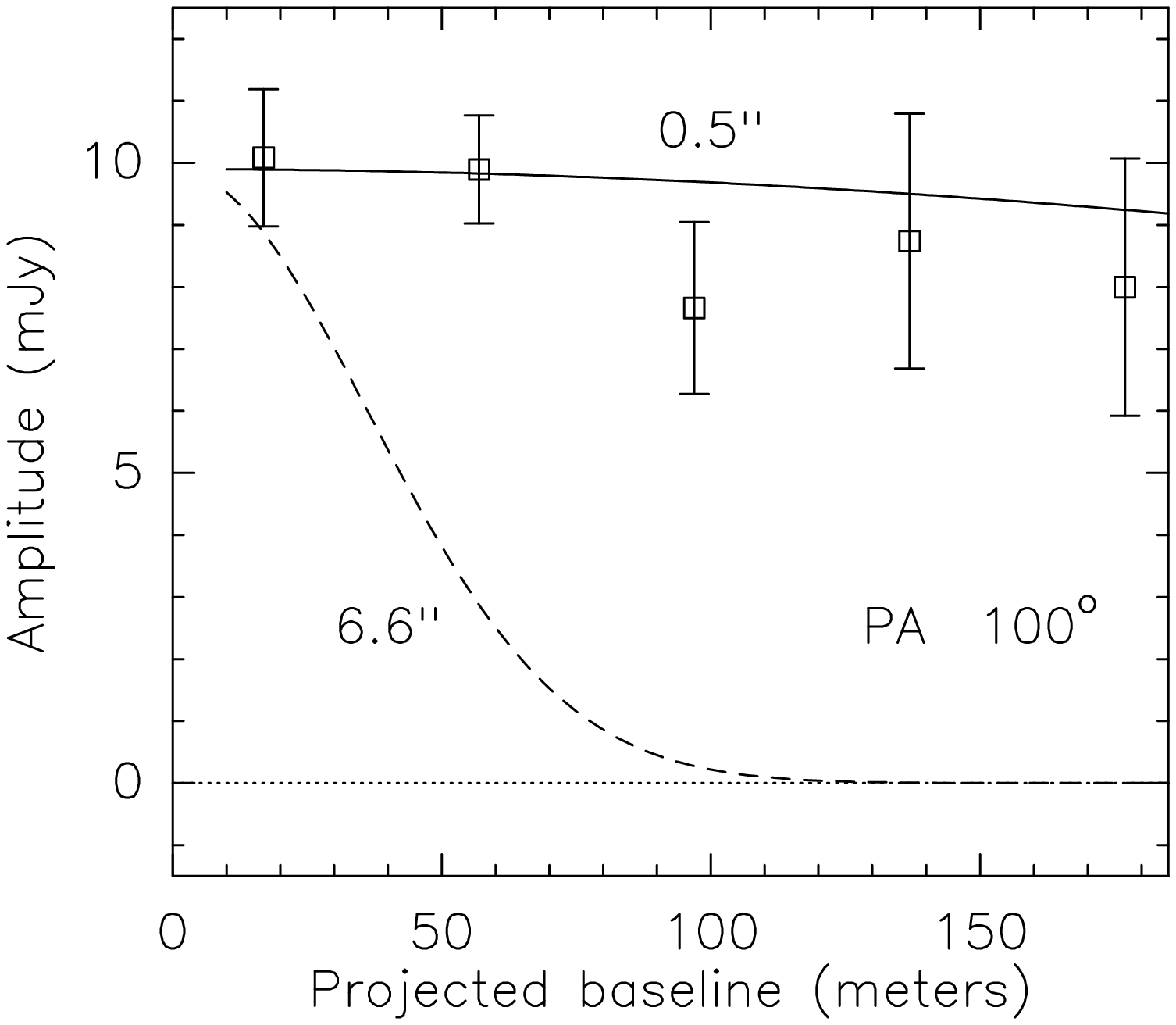}
\end{figure}
\begin{figure}
\vspace{-2.5cm} \caption[CO(3--2) visibilities north-south and
east-west.] { Size measurement: CO(3--2) visibility amplitudes
north-south and east-west.  The solid curves are the best fits for
an elliptical gaussian with half-power widths $2.2''\times \leq
0.5''$, at p.a.\ $10^\circ$ on the sky.  The dashed curves show
where the amplitude would have fallen if the source size had been
6.6$''$.  {\it Left:} Visibility $\sim$ north-south (p.a.\
$10^\circ$), averaging step 20\,m.

{\it Right:} Visibility $\sim$ east-west (p.a.\ $100^\circ$),
averaging step 40\,m.  For both figures, the CO profile was integrated
over 260\,\kms , centered on the CO peak of SMM J14011+0252. Error
bars are 1-$\sigma$.
}
\end{figure}
\clearpage
\newpage
%
\begin{figure}
\caption[CO(3--2) and (7--6) spectral grids]
{CO spectra with 20\,\kms\ resolution, on grids
around CO the peak of SMM J14011+0252.

{\it  Left:}
CO(3--2) spectra with a grid spacing of 2$''$.  The beam is
$5.9'' \times 4.5''$, with $T_b/S$ = 4.9\,K/Jy, and the r.m.s.\ noise is
0.8\,mJy.  Velocity offsets are relative to 96.989\,GHz.
%

{ \it  Right:}
CO(7--6) spectra with a grid spacing of 1$''$.  The beam is $2.2''
\times 2.0''$ with $T_b/S$ = 5.4\,K/Jy, and the r.m.s.\ noise is
3.1\,mJy.  Velocity offsets are relative to 226.251\,GHz.
}\end{figure}
%
\begin{figure}
\caption[CO(3--2) and (7--6) channel maps.]
{CO channel maps of SMM J14011+0252.

{\it  Left:} CO(3--2) in 20\,\kms\ channels with a $5.9'' \times
4.5''$ beam (lower right box), with $T_b/S$ = 4.9\,K/Jy. Contour
steps are 1.6\,mJy ($2\sigma$). Velocity offsets (upper left
corners) are relative to 96.989\,GHz.

{\it Right:} CO(7--6) in 40\,\kms\ channels with a $2.2'' \times
2.0''$ beam (lower right box), with $T_b/S$ = 5.4\,K/Jy. Contours
start at 4.4\,mJy ($2\sigma$), and increase in steps of 2.2\,mJy
($1\sigma$).  Velocity offsets (upper left corners) are relative
to 226.251\,GHz.  In all diagrams, the cross marks the initial CO
position from Frayer et al.\ (1999). }\end{figure} \clearpage
%
\begin{figure}
\vspace{-2.0cm}
\caption[HST images of 14011+0252 and 0414+0534]
{Comparison of {\it HST} images of SMM J14011+0252 and
MG 0414+0534.

{\it  Left: }
{\it HST R}-band field of SMM J14011+0252, covering 6.0$''$ on a side,
with tick marks every 1$''$ (from Ivison et al.\ 2001).  The labels give
our interpretation of the objects as lensing galaxy, close double
J1-southeast, and two counter-images.

{\it  Right: } {\it HST I}-band exposure of MG 0414+0534, covering
5.8$''$ on a side (from Falco et al.\ 1997).  We mark the lensing
galaxy, the close double A1+A2 and the two counter-images B and C.
In the model of Falco et al., the four spots are the quasar, while
the arc is a distorted view of a much larger ionized region, about
200\,pc from the quasar.

}\end{figure}

\clearpage
\begin{figure}
\vspace{2.0cm}
\caption[Gravitational lens model]
{A model that may explain  SMM J14011+0252.

{\it Left: Source plane.}
The lensing galaxy is at (0,0), the radial caustic (outer ellipse)
and the tangential caustic (inner diamond) are
pulled off-center by the mass of the $z=0.25$ cluster Abell 1835,
which is centered at ($-40''$, $+20''$).  The background ultraluminous
galaxy at $z=2.565$ has a compact UV core with a diameter of
0.02$''$ (black core), surrounded by a larger ring of molecular gas with a
diameter of 0.18$''$ (grey halo).

{\it Right: Lens Plane.}
The lensing galaxy is at (0,0), the outer ellipse is the tangential
critical curve and the inner ellipse is the radial critical curve,
corresponding to the caustics in the source plane.  The core UV light
of the background galaxy is deflected into small arclets (the black
central parts of the arcs), similar to the J1-southeast double and the
J2 and J1n components in the {\it HST R}-band field of 14011+0252.
The millimeter CO and dust radiation is distorted into the larger grey
arcs.  Most of the CO and dust flux is in the large arc, associated
with the J1 component.  }\end{figure}

\clearpage
\end{document}